\documentclass[%
reprint,
nofootinbib,
amsmath,amssymb,
aps,
]{revtex4-2}
\pdfoutput=1

\usepackage[margin=1.64cm]{geometry}
\usepackage{dcolumn}
\usepackage{bm}
\usepackage{siunitx}
\usepackage{braket}
\usepackage{rotating}

\usepackage{xcolor}
\usepackage{graphicx}
\usepackage{float}

\usepackage{booktabs}
\usepackage{tabularx}
\usepackage{listings}
\definecolor{mgreen}{HTML}{009901}
\definecolor{light-gray}{gray}{0.95}
\usepackage{xfrac}

\usepackage{hyperref}
\hypersetup{%
    hypertexnames=false,
    hidelinks
}
\usepackage[all]{hypcap} 
\usepackage[capitalise]{cleveref}

\usepackage[inline]{enumitem}

\usepackage[english]{babel}

\usepackage[utf8]{inputenc}
\usepackage[T1]{fontenc}

\usepackage[nomain,acronym,nonumberlist,style=list]{glossaries}
\newacronym{NV}{NV}{Nitrogen-Vacancy}
\newacronym{AWG}{AWG}{Arbitrary Waveform Generator}
\newacronym{QEGP}{QEGP}{Quantum Entanglement Generation Protocol}
\newacronym{MHP}{MHP}{Midpoint Heralding Protocol}
\newacronym{TDMA}{TDMA}{Time Division Multiple Access}
\newacronym{SPI}{SPI}{Serial Peripheral Interface}
\newacronym{ENT}{ENT}{Entanglement attempts command (\textit{K}-type)}
\newacronym{ENM}{ENM}{Entanglement attempts command (\textit{M}-type)}
\newacronym{SQG}{SQG}{Single Qubit Gate command}
\newacronym{INI}{INI}{Qubit Initialization command}
\newacronym{MSR}{MSR}{Qubit Measurement command}
\newacronym{PMG}{PMG}{Pre-Measurement Gate command}
\newacronym{HAL}{HAL}{Hardware Abstraction Layer}
\newacronym{tcpip}{TCP/IP}{Transmission Control Protocol / Internet Protocol. Also known as the Internet Protocol suite}
\newacronym{dqp}{DQP}{Distributed queue protocol }
\makenoidxglossaries

\begin{document}

\title{Experimental demonstration of entanglement delivery using a quantum network stack}

\author{M. Pompili}\altaffiliation{These authors contributed equally to this work}
\author{C. Delle Donne}\altaffiliation{These authors contributed equally to this work}
\affiliation{%
\vspace{0.5em}QuTech \& Kavli Institute of Nanoscience, Delft University of Technology, 2628 CJ Delft, The Netherlands
}

\author{I. te Raa}
\author{B. van der Vecht}
\author{M. Skrzypczyk}
\author{G. Ferreira}
\author{L. de Kluijver}
\author{A. J. Stolk}
\author{S. L. N. Hermans}
\author{P. Pawe\l{}czak}
\author{W. Kozlowski}
\author{R. Hanson}\email[Correspondence to: ]{R.Hanson@tudelft.nl}
\author{S. Wehner}\email[Correspondence to: ]{S.D.C.Wehner@tudelft.nl}
\affiliation{%
\vspace{0.5em}QuTech \& Kavli Institute of Nanoscience, Delft University of Technology, 2628 CJ Delft, The Netherlands
}

\begin{abstract}
Scaling current quantum communication demonstrations to a large-scale quantum network will require not only advancements in quantum hardware capabilities, but also robust control of such devices to bridge the gap to user demand. Moreover, the abstraction of tasks and services offered by the quantum network should enable platform-independent applications to be executed without knowledge of the underlying physical implementation. Here we experimentally demonstrate, using remote solid-state quantum network nodes, a link layer and a physical layer protocol for entanglement-based quantum networks. The link layer abstracts the physical-layer entanglement attempts into a robust, platform-independent entanglement delivery service. The system is used to run full state tomography of the delivered entangled states, as well as preparation of a remote qubit state on a server by its client. Our results mark a clear transition from physics experiments to quantum communication systems, which will enable the development and testing of components of future quantum networks.
\end{abstract}

\maketitle

\section{Introduction}

By sharing entangled states over large distances, the future Quantum Internet~\cite{kimble_quantum_2008, wehner_quantum_2018} can unlock new possibilities in secure communication~\cite{ekert_ultimate_2014}, distributed and blind quantum computation~\cite{jiang_distributed_2007, broadbent_universal_2009}, and metrology~\cite{gottesman_longer-baseline_2012, komar_quantum_2014}.
Fundamental primitives for entanglement-based quantum networks have been demonstrated across several physical platforms, including trapped ions~\cite{moehring_entanglement_2007, stephenson_high-rate_2020}, neutral atoms~\cite{ritter_elementary_2012, hofmann_heralded_2012}, diamond color centers~\cite{bernien_heralded_2013, kalb_entanglement_2017, humphreys_deterministic_2018, pompili_realization_2021},
and quantum dots~\cite{delteil_generation_2016, stockill_phase-tuned_2017}.
To scale up such physics experiments to intermediate-scale quantum networks, researchers have been investigating how to enclose the complex nature of quantum entanglement generation into more robust abstractions~\cite{aparicio_protocol_2011, dahlberg_link_2019, pirker_quantum_2019, kozlowski_designing_2020, aguado_enabling_2020, kozlowski_p4_2020, alshowkan_reconfigurable_2021}.

\begin{figure}
    \centering
    \includegraphics{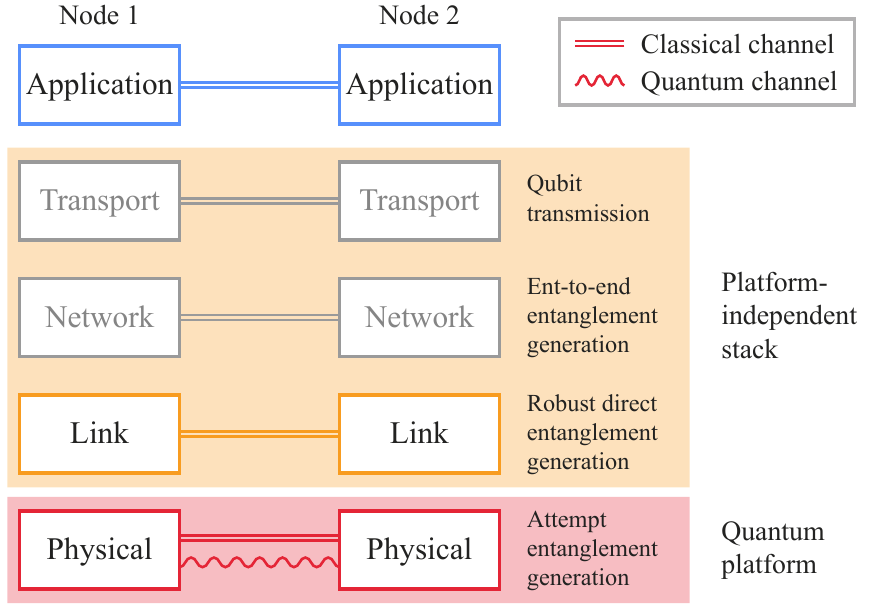}
    \caption{Quantum network stack architecture.
    At the bottom of the stack, the physical layer (red), which is highly quantum platform-dependent, is tasked with attempting entanglement generation.
    The link layer (yellow) uses the functionality provided by the physical layer to provide a platform-independent and robust entanglement generation service between neighboring nodes to the higher layers.
    Network and transport layer (not implemented in this work, grayed out) will support end-to-end connectivity and qubit transmission.
    Applications (blue) use the services offered by the stack to perform quantum networking tasks. Based on~\citet{dahlberg_link_2019}.}
    \label{fig:link_figure_1}
\end{figure}

A natural way to render a complex system scalable, is to design its architecture as a stack of layers that go from specialized physical medium protocols to more general services.
Inspired by the TCP/IP protocol stack commonly employed in classical networks, similar stacks have been proposed for quantum networks~\cite{dahlberg_link_2019, pirker_quantum_2019, kozlowski_designing_2020}, like the one depicted in \Cref{fig:link_figure_1}.
These recent efforts, along with proposals for resource scheduling and routing techniques (e.g.~\cite{van_meter_path_2013, caleffi_optimal_2017, gyongyosi_decentralized_2018, pant_routing_2019, chakraborty_distributed_2019, shi_concurrent_2020, chakraborty_entanglement_2020}), pave the way for larger-scale quantum networks.

In this work we experimentally demonstrate---for the first time---a link layer protocol for entanglement-based quantum networks.
The link layer abstracts the generation of entangled states between two physically separated solid-state qubits into a robust and platform-independent service.
An application can request entangled states from the link layer and then, in addition, apply local quantum operations on the entangled qubits in real-time.
Using the link layer, we perform full state tomography of the generated states and achieve remote state preparation---a building block for blind quantum computation---as well as measuring the latency of the entanglement generation service.

To evaluate correct operation and performance of our system, we measure
\begin{enumerate*}[label=(\alph*)]
    \item the fidelity of the generated states and
    \item the latency incurred by link layer and physical layer when generating entangled pairs.
\end{enumerate*}
For both fidelity and latency, we find that our system performs with marginal overhead with respect to previous non-platform-independent experiments. We also identify the sources of the additional overhead incurred, and propose improvements for future realizations.

\section{Quantum Link Layer Protocol}
\label{sec:link_layer}

Remote entanglement generation constitutes a fundamental building block of quantum networking.
However, for a user to be able to integrate it into more complex quantum networking applications and protocols, the entanglement generation service must also be:
\begin{enumerate*}[label=(\alph*)]
    \item robust, meaning that the user should not have to deal with entanglement failures and retries, and that an entanglement request should result in the delivery of an entangled pair;
    \item quantum platform-independent, in order for the user to be able to request entanglement without having to understand the inner workings of the underlying physical implementation;
    \item on-demand, such that the user can request and consume entanglement as part of a larger quantum communication application.
\end{enumerate*}
Robust, platform-independent, on-demand entanglement generation must figure as one of the basic services offered by a system running on a quantum network node.
In other words, establishing a reliable quantum link between two directly connected nodes is the task of the first layer above the physical layer in a quantum networking protocol stack, as portrayed in \Cref{fig:link_figure_1}.
Following the TCP/IP (Internet protocol suite) model nomenclature, we refer to this layer as the \emph{link layer}.
We remark that, in the framework of a multi-node network, a quantum network stack should also feature a \emph{network layer} (called \emph{internet layer} in the TCP/IP model) to establish links between non-adjacent nodes, and optionally a \emph{transport layer} to encapsulate qubit transmission into a service~\cite{dahlberg_link_2019,kozlowski_designing_2020,pirker_quantum_2019} (as shown in \cref{fig:link_figure_1}).

\subsection{Link Layer Service}

The service provided by a link layer protocol for quantum networks should expose a few configuration parameters to its user.
To ensure a platform-independent interaction with the link layer, such parameters should be common to all possible implementations of the quantum physical device.
In this work, we implement a revised version of the link layer protocol proposed---but not implemented---in Ref.~\cite{dahlberg_link_2019}, with the following service description. 
The interface exposed by the link layer should allow the higher layer to specify:
\begin{enumerate*}[label=(\alph*)]
    \item \emph{Remote node ID}, an identifier of the remote node to produce entanglement with (in case the requesting node has multiple neighbors);
    \item \emph{Number of entangled pairs}, to allow for the creation of several pairs with one request;
    \item \emph{Minimum fidelity}, an indication of the desired minimum fidelity for the produced pairs;
    \item \emph{Delivery type}, whether to keep the produced pair for future use (type \textit{K}), measure it directly after creation (type \textit{M}), or measure the local qubit immediately and instruct the remote node to keep its own for future use (type \textit{R}, used for remote state preparation);
    \item \emph{Measurement basis}, the basis to use when measuring \textit{M}- or \textit{R}-type entangled pairs;
    \item \emph{Request timeout}, to indicate a time limit for the processing of the request.
\end{enumerate*}
After submitting an entanglement generation request, the user should expect the link layer to coordinate with the remote node and to handle entanglement generation attempts and retries until all the desired pairs are produced (or until the timeout has expired).
When completing an entanglement generation request, the link layer should then report to the above layer the following:
\begin{enumerate*}[label=(\alph*)]
    \item \emph{Produced Bell state}, the result of entanglement generation;
    \item \emph{Measurement outcome}, in case of \textit{M}- or \textit{R}-type entanglement requests;
    \item \emph{Entanglement ID}, to uniquely identify an entangled pair consistently across source and destination of the request.
\end{enumerate*}

\subsection{A Quantum Link Layer Protocol}

A design of a quantum link layer protocol that offers the above service is the \emph{quantum entanglement generation protocol} (QEGP) proposed by \citet{dahlberg_link_2019}.
As originally designed, this protocol relies on the underlying quantum physical layer protocol to achieve accurate timing synchronization with its remote peer and to detect inconsistencies between the local state and the state of the remote counterpart.
To satisfy such requirements, QEGP is accompanied by a quantum physical layer protocol, called \emph{midpoint heralding protocol} (MHP), designed to support QEGP on heralded entanglement-based quantum links.

\textbf{Entanglement requests and agreement.}
QEGP exposes an interface for its user to submit \emph{entanglement requests}.
An entanglement request can specify all the aforementioned configuration parameters (remote node ID, number of entangled pairs, minimum fidelity, request type, measurement basis), and an additional set of parameters which can be used to determine the priority of the request.
In the theoretical protocol proposed in Ref.~\cite{dahlberg_link_2019}, agreement on the requests between the nodes is achieved using a distributed queue protocol (DQP) which adds the incoming requests to a joint queue.
The distributed queue, managed by the node designated as primary, ensures that both nodes schedule pending entanglement requests in the same order.
Moreover, QEGP attaches a timestamp to each request in the distributed queue, so that both nodes can process the same entanglement request simultaneously.

\textbf{Time synchronization.}
Time-scheduling entanglement generation requests is necessary for the two neighboring nodes to trigger entanglement generation at the same time, and avoid wasting entanglement attempts.
QEGP relies on MHP to maintain and distribute a synchronized clock, which QEGP itself uses to schedule entanglement requests.
The granularity of such a clock is only marginally important, but its consistency across the two neighboring nodes is paramount to make sure that entanglement attempts are triggered simultaneously on the two ends.

\textbf{Mismatch verification.}
One of the main responsibilities of MHP is to verify that both nodes involved in entanglement generation are servicing the same QEGP request at the same time, which the protocol achieves by sending an auxiliary classical message to the heralding station when the physical device sends the flying qubit.
The heralding station can thus verify that the messages fetched by the two MHP peers are consistent and correspond to the same QEGP request.

\textbf{QEGP challenges.}
We identify three main challenges that would be faced when deploying QEGP on a large-scale quantum network, while suggesting an alternative solution for each of these.
\begin{enumerate*}[label=(C\arabic*)]
    \item \label{enum:link_ch_queue} Using a link-local protocol (DQP) to schedule entanglement requests, albeit sufficient for a single-link network, becomes challenging in larger networks, given that a node might be connected to more than just one peer.
    In such scenarios, the scheduling of entanglement requests can instead be deferred to a centralized scheduling entity, one which has more comprehensive knowledge of the entire (sub)network~\cite{skrzypczyk_dynamic_2021}.
    \item \label{enum:link_ch_tsync} Entrusting the triggering of entanglement attempts to QEGP would impose very stringent real-time constraints on the system where QEGP itself is deployed---even microsecond-level latencies on either side of the link can result in out-of-sync (thus wasteful) entanglement attempts.
    While \citet{dahlberg_link_2019} identify this problem as well, the original MHP protocol assumes that both QEGP peers issue an entanglement command to the physical layer at the same clock cycle.
    In this scheme, MHP initiates an entanglement attempt regardless of the state of the remote counterpart.
    We believe that fine-grained entanglement attempt synchronization should pertain to the physical layer only, building on the assumption that the real-time controllers deployed at the physical layer of each node are anyway highly synchronized~\cite{pompili_realization_2021}.
    \item \label{enum:link_ch_mismatch} Checking for request mismatches at the heralding station requires the latter to be capable of performing such checks in real-time.
    Given that the two neighboring MHP protocols have to anyway synchronize before attempting entanglement, we suggest that, as an alternative approach, consistency checks be performed at the nodes themselves, rather than at the heralding station, just before entering the entanglement attempt routine.
\end{enumerate*}

\subsection{Revised Protocol}

To address the present QEGP and MHP challenges with the proposed solutions, we have made some modifications to the original design of the two protocols.
In particular, we adopted a centralized request scheduling mechanism~\cite{skrzypczyk_dynamic_2021} to tackle challenge~\ref{enum:link_ch_queue}, we delegated the ultimate triggering of entanglement attempts to MHP as a solution to challenge~\ref{enum:link_ch_tsync}, and we assigned request mismatch verification to the MHP protocol running on each node, rather than to the heralding station, to address challenge~\ref{enum:link_ch_mismatch}.

\textbf{Centralized request scheduling.}
To avoid using a link-local protocol (DQP) to schedule entanglement requests, our version of QEGP defers request scheduling to a \emph{centralized request scheduler}, whereby a node's entanglement generation schedule is computed on the basis of the whole network's needs.
Delegating network scheduling jobs to centralized entities is, albeit not the only alternative, a common paradigm of classical networks, and especially of software-defined networking (SDN)---a concept that has been recently investigated in the context of quantum networking~\cite{aguado_enabling_2020,kozlowski_p4_2020}.
In our system, the centralized scheduler produces a time-division multiple access (TDMA) network schedule---one for each node in the network---where each time bin is reserved for a certain class of entanglement generation requests~\cite{skrzypczyk_dynamic_2021}.
A class of requests may comprise, for instance, all requests coming from the same application and asking for the same fidelity of the entangled states.
While reserving time bins may be redundant in a single-link network, integrating a centralized scheduling mechanism early on into the link layer protocol will facilitate future developments.

\textbf{MHP synchronization and timeout.}
Although centralized request scheduling makes the synchronization of QEGP peers easier, precise triggering of entanglement attempts should still be entrusted to the component of the system where time is the most deterministic---in our case, the physical layer protocol MHP.
In contrast to Ref.~\cite{dahlberg_link_2019}, once MHP fetches an entanglement instruction from QEGP, the protocol announces itself as ready to its remote peer, and waits for the latter to do so as well.
After this synchronization step succeeds, the two MHP peers can instruct the underlying hardware to trigger an entanglement attempt at a precise point in time.
If, instead, one of the two MHP peers does not receive announcements from its remote counterpart within a set timeout, it can conclude that the latter is not ready, or temporarily not responsive, and can thus return control to QEGP without wasting entanglement attempts.
This MHP synchronization step is also useful for the two sides to verify that they are processing the same QEGP request, and thus catch mismatches.

The MHP synchronization routine inherently incurs some overhead, which is also larger on longer links.
We mitigate this overhead by batching entanglement attempts---that is, the physical layer attempts entanglement multiple times after synchronization before reporting back to the link layer.
The maximum number of attempts per batch is a purely physical-layer parameter, and it has no relation with the link layer entanglement request timeout parameter described in Ref.~\cite{dahlberg_link_2019}---although batches should be small enough for the link layer timeout to make sense.

\subsection{Implementation}

\begin{figure}
    \centering
    \includegraphics{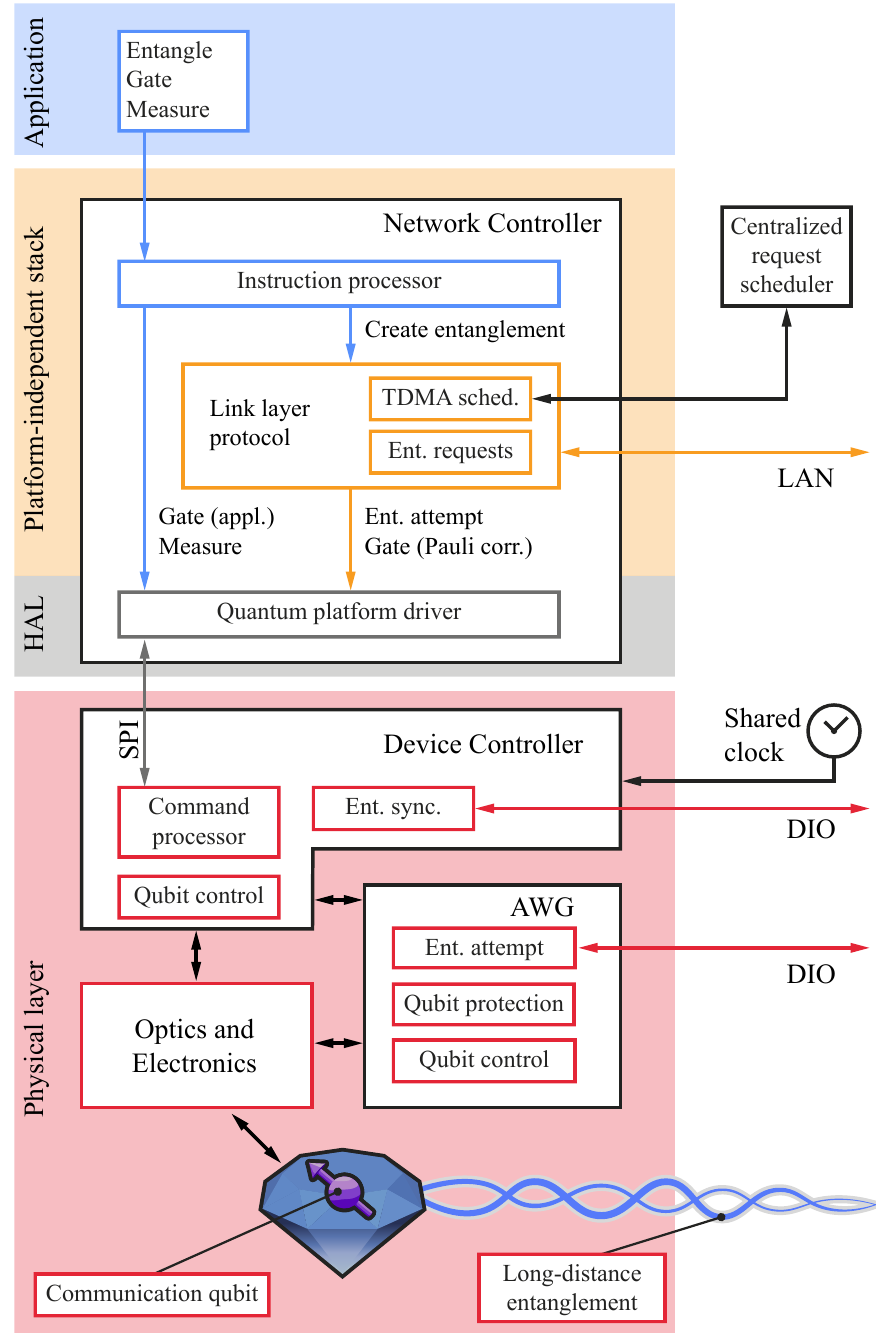}
    \caption{Quantum network node architecture.
    From top to bottom:
    At the application layer, a simple platform-independent routine is sent to the network controller.
    The network controller implements the platform-independent stack---in this work only the link layer protocol---and a hardware abstraction layer (HAL) to interface with the physical layer's device controller.
    An instruction processor dispatches instructions either directly to the physical layer, or to the link layer protocol in case a remote entangled state is requested by the application.
    The link layer schedules entanglement requests and synchronizes with the remote node (on a local area network, LAN) using a time-division multiple access (TDMA) schedule computed by a centralized scheduler (external).
    At the physical layer, the device controller fetches commands from---and replies with outcomes to---the network controller.
    Driven by a clock shared with the neighboring node, it performs hard-real-time synchronization for entanglement generation using a digital input/output (DIO) interface.
    By controlling the optical and electronic components (among which an arbitrary waveform generator, AWG), the device controller can perform universal quantum control of the communication qubit in real-time, as well as attempt long-distance entanglement generation with the neighboring node.
    }
    \label{fig:link_figure_2}
\end{figure}

The original design of the QEGP and MHP protocols, as well as our revision, specifies the conceptual interaction between the two protocols and the service exposed to a higher layer in the system, but does not impose particular constraints on how to implement link layer and physical layer, how to realize the physical interface between them, and how to configure things such as the centralized request scheduler and the entanglement attempt procedure.
\Cref{fig:link_figure_2} gives an overview of the architecture of our quantum network nodes.
We briefly describe our most relevant implementation choices here and in \Cref{sec:link_physical_layer}.

\textbf{Application processing.}
At the application layer, user programs---written in Python using a dedicated \emph{software development kit}~\cite{netqasm_sdk}---are processed by a rudimentary compilation stage, which translates abstract quantum networking applications into gates and operations supported by our specific quantum physical platform.
Such gates and operations are expressed in a low-level assembly-like language for quantum networking applications called \emph{NetQASM}~\cite{dahlberg_netqasm_2021}.
As part of our software stack, we also include an \emph{instruction processor}, conceptually placed above the link layer, which is in charge of dispatching entanglement requests to QEGP and other application instructions to the physical layer directly.

\textbf{Interface.}
Ref.~\cite{dahlberg_link_2019} did not provide a specification of the interface to be exposed by the physical layer.
We designed this interface such that the physical layer can accept \emph{commands} from the higher layer, specifically:
\begin{enumerate*}[label=(\alph*)]
    \item qubit initialization (\texttt{INI}),
    \item qubit measurement (\texttt{MSR}),
    \item single-qubit gate (\texttt{SQG}),
    \item entanglement attempt (\texttt{ENT}, or \texttt{ENM} for \textit{M}- or \textit{R}-type requests),
    \item pre-measurement gates selection (\texttt{PMG}, to specify in which basis to measure the qubit for \textit{M}- or \textit{R}-type requests).
\end{enumerate*}
For each command, the physical layer reports back an \emph{outcome}, which indicates whether the command was executed correctly, and can bear the result of a qubit measurement and the Bell state produced after a successful entanglement attempt.
Our software stack also comprises a \emph{hardware abstraction layer} (HAL) that sits below QEGP and the instruction processor.
The HAL encodes and serializes commands and outcomes, and is thus used to interface with the device controller.

\textbf{TDMA network schedule.}
Designing a full-blown centralized request scheduler is a challenge in and of its own, outside the scope of this work.
Instead of implementing such a scheduler, we compute static TDMA network schedules~\cite{skrzypczyk_dynamic_2021} and install them manually on the two network nodes upon initialization.
TDMA schedules for our simple single-link experiments are quite trivial (see \cref{sec:link_suppl_tdma}), as the network resources of a node are not contended by multiple links.

\textbf{Entanglement attempts.}
Producing entanglement on a link can take several attempts.
To minimize the number of \texttt{ENT} commands fetched by MHP from QEGP, as well as to mitigate the MHP synchronization overhead incurred after each entanglement command, we batch entanglement attempts at the MHP layer, such that synchronization and outcome reporting only happens once per batch of attempts.

\textbf{Delivered entangled states.}
In our first iteration, we implemented QEGP such that it always delivers $\ket{\Phi^+}$ Bell states to the higher layer.
This means that, when the physical layer produces a different Bell state, QEGP (on the node where the entanglement request originates) issues a single-qubit gate---a Pauli correction---to transform the entangled pair into the $\ket{\Phi^+}$ state\footnote{We abbreviate the four two-qubit maximally entangled Bell states as $\ket{\Phi^\pm} = (\ket{00} \pm \ket{11})/\sqrt{2}$ and $\ket{\Psi^\pm} = (\ket{01} \pm \ket{10})/\sqrt{2}$.}.
A future version of QEGP could allow the user to request any Bell state, and could extract the Pauli correction from QEGP so that the application itself can decide, depending on the use case, whether to apply the correction or not.

\textbf{Mismatch verification.}
As per our design specification, MHP should also be responsible for verifying that the entanglement commands coming from the two QEGP peers belong to the same request.
We did not implement this feature yet because, in our simple quantum network, we do not expect losses on the classical channel used by the two MHP parties to communicate---a lossy classical channel would be the primary source of inconsistencies at the MHP layer~\cite{dahlberg_link_2019}.
However, we believe that this verification step will prove very useful in real-world networks where classical channels do not behave as predictably.

\textbf{Deployment.}
We implemented QEGP as a software module in a system that also includes the instruction processor and the hardware abstraction layer.
QEGP, the instruction processor and the hardware abstraction layer, forming the \emph{network controller}, are implemented as a C/C++ standalone runtime developed on top of FreeRTOS, a real-time operating system for embedded platforms~\cite{freertos}.
The runtime and the underlying operating system are deployed on a dedicated Avnet MicroZed---an off-the-shelf platform based on the Zynq-7000 SoC, which hosts two ARM Cortex-A9 processing cores, of which only one is used, clocked at \SI{667}{\MHz}.
QEGP connects to its remote peer via TCP over a Gigabit Ethernet interface.
The interface to the physical layer is realized through a \SI{12.5}{\MHz} SPI connection.
The user application is sent from a general-purpose 4-core desktop machine running Linux, which connects to the instruction processor through the same Gigabit Ethernet interface that QEGP uses to communicate with its peer.
\section{Physical Layer Control in Real-Time}
\label{sec:link_physical_layer}

In this section, we outline the design and operation of the physical layer, which executes the commands issued by the higher layers on the quantum hardware and handles time-critical synchronization between the quantum network nodes.
The physical layer of a quantum network, as opposed to the apparatus of a physics experiment, needs to be able to execute commands coming from the layer above in real-time. Additionally, when performing the requested operations, it needs to leave the quantum device in a state that is compatible with future commands (for example, as discussed below, it should protect qubits from decoherence while it awaits further instructions). Finally, if a request cannot be met (e.g.~the local quantum hardware is not ready, the remote quantum hardware is not available, etc.), the physical layer should notify the link layer of the issue without interrupting its service.

Our quantum network is composed of two independent nodes based on diamond NV centers physically separated by \SI{\approx 2}{m} (see \cref{fig:link_figure_2} for the architecture of one node, and \cref{fig:link_suppl_setup} for details on the connections between the two nodes). We will refer to the two nodes as \textit{client} and \textit{server}, noting that this is only a logical separation useful to describe the case studies---the two nodes have the exact same capabilities.
On each node, we implement the logic of the physical layer in a state-machine-based algorithm deployed on a time-deterministic microcontroller, the \emph{device controller} (J\"ager ADwin Pro II, based on Zynq-7000 SoC, dual-core ARM Cortex-A9, clocked at \SI{1}{\GHz}). Additionally, each node uses an arbitrary waveform generator (AWG, Zurich Instruments HDAWG8, \SI{2.4}{GSa/s}, \SI{300}{\MHz} sequencer) for nanosecond-resolution tasks, such as fast optical and electrical pulses; the use of such a user-programmable FPGA-based AWG, as opposed to a more traditional upload-and-play instrument (such as the ones used in Ref.~\cite{pompili_realization_2021}), enables the real-time control of our quantum device.

\subsection{Single node operation}

On our quantum platform, before a node is available to execute commands, it needs to perform a qubit readiness procedure called \emph{charge and resonance check} (CR check). This ensures that the qubit system is in the correct charge state and that the necessary lasers are resonant with their respective optical transitions. Other quantum platforms might have a similar preparation step, such as loading and cooling for atoms and ions~\cite{stephenson_high-rate_2020, ritter_elementary_2012}.
Once the CR check is successful, the device controller can fetch a command from the network controller. Depending on the nature of the command, the device controller might need to coordinate with other equipment in the node or synchronize with the device controller of the other node.

For qubit initialization and measurement commands (\texttt{INI} and \texttt{MSR}), the device controller shines the appropriate laser for a pre-defined duration (\texttt{INI}\SI{\approx100}{\us}, \texttt{MSR}\SI{\approx10}{\us}). Both operations are deterministic and carried out entirely by the device controller.

Single qubit gates (\texttt{SQG}) require the coordination of the device controller and the AWG. For our communication qubits, they consist of generating an electrical pulse with the AWG (duration \SI{\approx100}{\ns}), which is then multiplied to the qubit frequency (\SI{\approx2}{\GHz}), amplified and finally delivered to the quantum device.
The link layer can request rotations in steps of $\pi/16$ around the X, Y or Z axis of the Bloch sphere (here we implement only X and Y rotations, Z rotations will be implemented in the near future, see \cref{sec:link_suppl_sqg}).
When a new gate is requested by the link layer, the device controller at the physical layer informs the AWG of the gate request via a parallel \num{32}-bit DIO interface. The AWG will then select one of the $64$ pre-compiled waveforms, play it, and notify the device controller that the gate has been executed. The device controller will in turn notify the network controller of the successful operation.

After the rotation has been performed, our qubit---if left idling---would lose coherence in \SI{\approx5}{\us}.
A coherence time exceeding \SI{1}{s} has been reported on our platform~\cite{abobeih_one-second_2018} using decoupling sequences (periodic rotations of the qubit that shield it from environmental noise). By interleaving decoupling sequences and gates, one can perform extended quantum computations~\cite{bradley_ten-qubit_2019}. These long sequences of pulses have in the past been calculated and optimized offline (on a PC), then uploaded to an AWG, and finally executed on the quantum devices with minimal interaction capabilities (mostly binary branching trees, see~\cite{pompili_realization_2021}).
In our case, it is impossible to pre-calculate these sequences, since we cannot know in advance which gates are going to be requested by the link layer.
To solve this challenge, we implement a \textit{qubit protection} module on the AWG, that interleaves decoupling sequences with the requested gates in real-time.
As soon as the first gate in a sequence is requested, the AWG starts a decoupling sequence on the qubit. Then, it periodically checks if a new gate has been requested, and if so, it plays it at the right time in the decoupling sequence. The AWG will continue the qubit protection routine until the device controller will ask for it to stop (e.g.~to perform a measurement). This technique allows us to execute universal qubit control without prior knowledge of the sequence to be played, and---crucially---in real-time.

\subsection{Entanglement generation}

Differently from the commands previously discussed, attempting entanglement generation (\texttt{ENT}) requires tight timing synchronization between the device controllers---and AWGs---of the two nodes.
In our implementation, the two device controllers share a common \SI{1}{MHz} clock as well as a DIO connection to exchange synchronization messages (see Ref.~\cite{pompili_realization_2021}). When the device controllers are booted, they synchronize an internal cycle counter that is used for time-keeping, and is shared, at each node, with their respective network controllers to provide timing information to the link layer and the higher layers. Over larger distances, one could use well-established protocols to achieve sub-nanosecond, synchronized, GPS-disciplined common clocks~\cite{serrano_white_2013}.

When a device controller fetches an \texttt{ENT} command, it starts a three-way handshake procedure with the device controller of the other node. If the other node has also fetched an \texttt{ENT} command, they will synchronize and proceed with the entanglement generation procedure. If one of the two nodes is not available (e.g.~it is still trying to pass the CR check) the other node will time out, after \SI{0.5}{\ms}, and return an \textit{entanglement synchronization failure} (\texttt{ENT\_SYNC\_FAIL}) to its link layer.
The duration of the timeout is chosen such that is comparable with the average time taken by a node to pass the charge and resonance check (if correctly on resonance). This is to avoid unnecessary interactions between physical layer and link layer.
After the entanglement synchronization step, the device controllers proceed with an optical phase stabilization cycle~\cite{pompili_realization_2021}, and then the AWGs are triggered to attempt entanglement generation. In our implementation, one device controller (the server's) triggers both AWGs to achieve sub-nanosecond jitter between the two AWGs (see \cref{sec:link_suppl_timing} for a discussion on longer distance implementation). Each entanglement attempt lasts \SI{3.8}{\us}, and includes fast qubit initialization, communication-qubit to flying-qubit entanglement, and probabilistic entanglement swapping of the flying qubits~\cite{pompili_realization_2021}. The AWGs attempt entanglement up to \num{1000} times before timing out and reporting an \textit{entanglement failure} (\texttt{ENT\_FAIL}). Longer batches of entanglement attempts would increase the probability that one of the nodes goes into the unwanted charge state (and therefore cannot produce entanglement, see \cref{sec:link_suppl_off_resonant}). While in principle possible, we did not implement, in this first realization, the charge stabilization mechanism proposed in Ref.~\cite{humphreys_deterministic_2018} that would allow for significantly longer batches of entanglement attempts.

If an entanglement generation attempt is successful (probability \num{\approx 5e-5}), the communication qubits of the two nodes will be projected into an entangled state (either $\ket{\Psi^+}$ or $\ket{\Psi^-}$, depending on which detector clicked at the heralding station).
To herald success of the entanglement attempt, a CPLD (Complex Programmable Logic Device, Altera MAX V 5M570ZF256C5N) sends a fast digital signal to both AWGs and device controllers, to prevent a new entanglement attempt from being played (which would destroy the generated entangled state). When the heralding signal is detected, the AWGs enter the qubit protection routine and wait for further instructions from the device controllers, which in turn notify the link layer of the successful entanglement generation, as well as which state was generated.

To satisfy \textit{M}- or \textit{R}-type entanglement requests, the link layer can instruct the physical layer to apply an immediate measurement to the entangled qubit by means of an \texttt{ENM} command.
Up until heralding of the entangled state, the physical layer operates as it does for the \texttt{ENT} command. When the state is ready, it proceeds immediately with a sequence of single qubit gates (as prescribed by an earlier \texttt{PMG} command) and a qubit measurement. The result of the measurement, together with which entangled state was generated, is communicated to the link layer. It is worth noting that the two nodes could fetch different types of requests and still generate entanglement. In fact, this will be used later in the remote state preparation application.
\section{Evaluation}
\label{sec:link_eval}

\begin{figure*}
    \centering
    \includegraphics[]{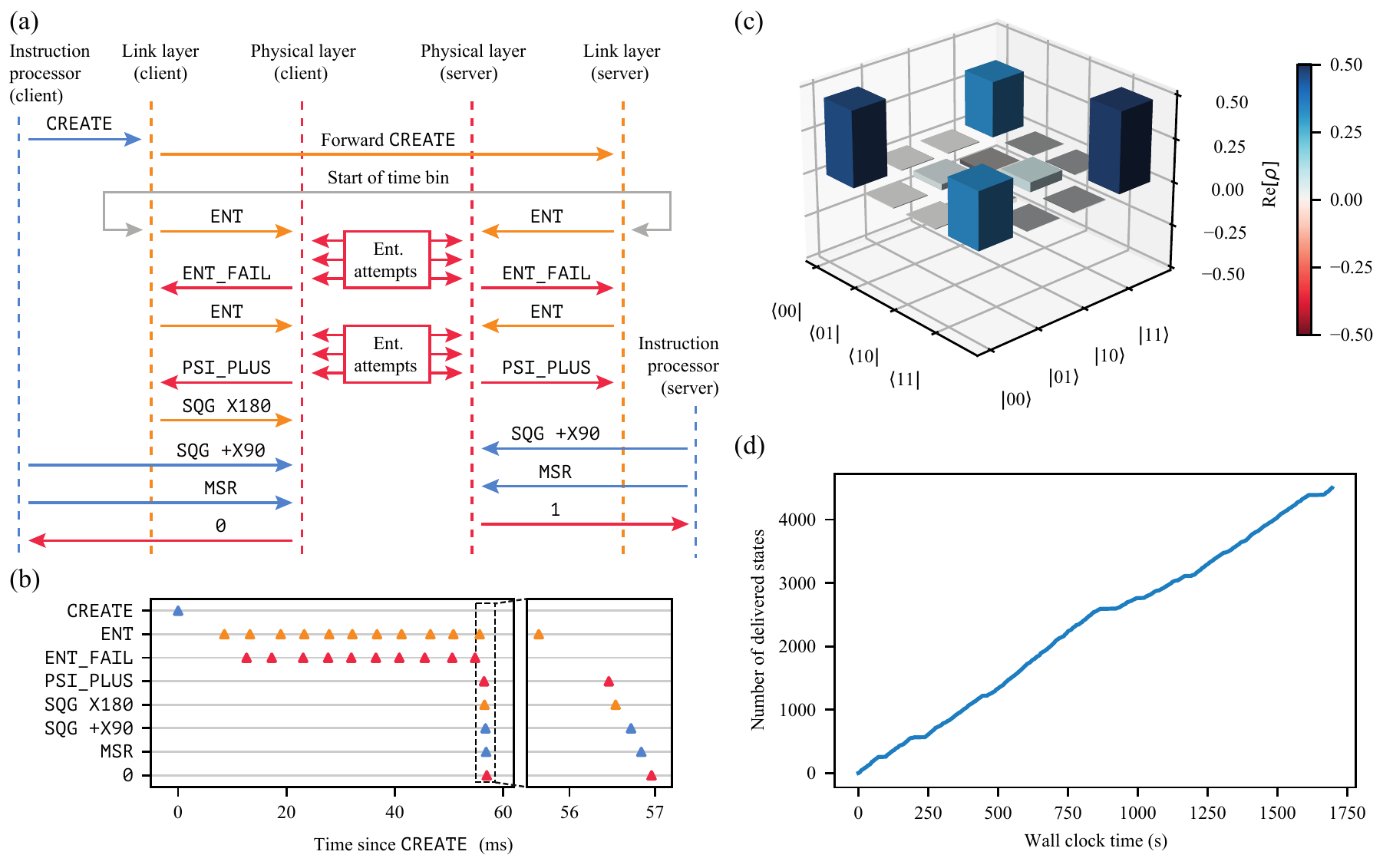}
    \caption{Full state tomography with the quantum network stack.
    \textbf{(a)} Sequence diagram of the communication steps across the network stack and the two nodes to perform one repetition of the tomography application (in particular, measurement of the $\braket{YY}$ correlator).
    The coloring follows that of \Cref{fig:link_figure_1}.
    \texttt{CREATE}: entanglement request, \texttt{ENT}: entanglement attempts request, \texttt{ENT\_FAIL}: failed the batch of entanglement attempts, \texttt{PSI\_PLUS}: successful entanglement attempt with generated state $\Psi^+$, \texttt{SQG}: single-qubit gate, \texttt{X180}: \ang{180} rotation around X axis, \texttt{MSR}: qubit measurement, \texttt{0/1}: qubit measurement outcome.
    See \Cref{tab:link_possible_outcomes} for a complete list of commands.
    Note that the client's link layer protocol requests a \texttt{X180} gate after entanglement generation to deliver the $\ket{\Phi^+}$ Bell state to the higher layer.
    \textbf{(b)} Example time trace of (a) for the client. Several batches of entanglement attempts are required before an entangled state is heralded. On the right, a zoomed-in part of the trace (corresponding to the dashed box in the left plot).
    \textbf{(c)} Reconstructed density matrix of the states delivered by the link layer.
    Only the real part is plotted (imaginary elements are all \num{\approx 0}, see main text).
    We estimate a fidelity F with $\ket{\Phi^+}$ of $F =$ \num{0.783(7)}.
    \textbf{(d)} Total number of delivered states over time.
    The occasional pauses in entanglement delivery (plateaus) are due to the client's NV center becoming off-resonant with the relevant lasers (see \cref{sec:link_suppl_off_resonant}).
    Differences in slope are due to changes in resonance conditions that increase the time necessary to pass the charge and resonance check.}
    \label{fig:link_figure_3}
\end{figure*}

To demonstrate and benchmark the capabilities of the link layer protocol, the physical layer, and of our system as a whole, we execute---on our two-node network---three quantum networking applications, all having a similar structure: the client asks for an entangled pair with the server, which QEGP delivers in the $\ket{\Phi^+}$ Bell state, and then both client and server measure their end of the pair in a certain basis.
First, we perform full quantum state tomography of the delivered entangled states.
Second, we request and characterize entangled states of varying fidelity.
Third, we execute remote preparation of qubit states on the server by the client.
For all three applications, we study the quality of the entangled pairs delivered by our system.
Additionally, we use the second application to assess the latency incurred by our link layer, and to compare it to the overall entanglement generation latency, including that of the physical layer.
Crucially, the three applications are executed back-to-back on the quantum network, without any software or hardware changes to the system---the only difference being the quantum-platform-independent application sent to the instruction processor (see~\ref{sec:link_suppl_apps}).

The sequence diagram in \Cref{fig:link_figure_3}a exemplifies the general flow between system components during the execution of an application.
At first, the instruction processor issues a request to create entanglement to link layer (\texttt{CREATE}).
Then, the client's link layer forwards the request with the server's counterpart (Forward \texttt{CREATE}).
The request is processed as soon as the designated time bin in the TDMA schedule starts, at which point the first entanglement command (\texttt{ENT}) is fetched by physical layer.
After an entangled state is produced successfully (\texttt{PSI\_PLUS}), the link layer of the client issues, if needed, a Pauli correction ($\pi$ rotation around the X axis, \texttt{SQG X180}) to deliver the pair in the $\ket{\Phi^+}$ state.
Finally, the instruction processor issues a gate ($\pi/2$ rotation around the X axis, \texttt{SQG X90}) and a measurement (\texttt{MSR}) to read out the entangled qubit in a certain basis, and receives an outcome from the physical layer (\texttt{0}).
\Cref{fig:link_figure_3}b illustrates the actual latencies between these interactions in one iteration of the full state tomography application.

For all our experiments, we configured TDMA time bins to be of \SI{20}{ms}.
In a larger network, the duration of time bins should be calibrated according to the average time it takes, on a certain link, to produce an entangled pair of a certain fidelity~\cite{skrzypczyk_dynamic_2021}.
By doing so, one can maximize network usage and thus reduce qubit decoherence on longer end-to-end paths.
However, in our single-link network, the duration of time bins only influences the frequency at which new entanglement requests are processed. Our time bin duration accommodates up to four batches of \num{1000} entanglement attempts.

\subsection{Full Quantum State Tomography}

The first application consists in generating entangled states at the highest \textit{minimum fidelity} currently available on our physical setup (\num{0.80}), and measuring the two entangled qubits in varying bases to learn their joint quantum state.
We measure all \num{9} two-node correlators ($\braket{\mathrm{XX}}, \braket{\mathrm{XY}}$, ...,~$\braket{\mathrm{ZZ}}$) as well as all their $\pm$ variations ($\braket{\mathrm{+X+X}}, \braket{\mathrm{+X-X}}$, etc.) to minimize the bias due to measurement errors.
For each of the $9 \times 4 = 36$ combinations, we measure \num{125} data points, for a total of \num{4500} entangled states generated and measured. \Cref{lst:link_fst} in \Cref{sec:link_suppl_apps} contains a pseudocode description of the application.

The collected measurement outcomes are then analyzed using QInfer~\cite{granade_qinfer_2017}, in particular the Monte Carlo method described in Ref.~\cite{granade_practical_2016} for Bayesian estimation of density matrices from tomographic measurements.
The reconstructed density matrix is displayed in \Cref{fig:link_figure_3}c (only the real part is shown in the figure) and its values and uncertainties are
\begin{equation*}
\mathrm{Re}[\rho] = \begin{pmatrix}
  0.442(6) & 0.003(3) & 0.003(2) & 0.328(5)\\
  0.003(3) & 0.033(6) & -0.023(5) & -0.000(5)\\
  0.003(2) & -0.023(5) & 0.056(4) & -0.003(4)\\
  0.328(5) & -0.000(5) & -0.003(4) & 0.469(7)\\
\end{pmatrix},
\end{equation*}
\begin{equation*}
\mathrm{Im}[\rho] = \begin{pmatrix}
  0 & -0.014(3) & -0.005(7) & 0.032(5)\\
  0.014(3) & 0 & -0.002(4) & 0.001(5)\\
  0.005(7) & 0.002(4) & 0 & -0.000(7)\\
  -0.032(5) & -0.001(5) & 0.000(7) & 0\\
\end{pmatrix}.
\end{equation*}
Here $\rho_{ij, mn} = \braket{ij|\ \rho\ |mn}$, with $i,m$ ($j,n$) being the client (server) qubit states in the computational basis.
The uncertainty on each element of the density matrix is calculated as the standard deviation of that element over the probability distribution approximated by the Monte Carlo reconstruction algorithm (probability distribution approximated by \num{1E5} Monte Carlo particles~\cite{granade_practical_2016}).
It is then possible to estimate the fidelity of the delivered entangled states with respect to the maximally entangled Bell state, which we find to be $F =$ \num{0.783(7)}.
The measured fidelity is slightly lower (\SI{\approx 3}{\percent}) than what measured in Ref.~\cite{pompili_realization_2021} without the use of the QEGP abstraction (and the whole network controller where QEGP runs).
This discrepancy could be due to the additional physical-layer decoupling sequences required for real-time operation (\SI{\approx 300}{\micro s}) and the additional single-qubit gate issued by the link layer to always deliver $\ket{\Phi^+}$ (see~\ref{sec:link_suppl_pauli_corr}).

It is to be noted that, in order to obtain the most faithful estimate of the generated state (see \cref{sec:link_suppl_m_corr} for details), the measured expectation values are corrected, in post-processing, to remove known tomography errors of both client and server~\cite{nachman_unfolding_2020}, and events in which at least one physical device was in the incorrect charge state.

Finally, we show, in \Cref{fig:link_figure_3}d, that our system can sustain a fairly stable entanglement delivery rate over \SI{\approx 30}{min} of data acquisition---plateaus and changes in slope can be attributed to varying conditions of resonance between the NV centers and the relevant lasers (see \cref{sec:link_suppl_off_resonant}).

\begin{figure}
    \centering
    \includegraphics{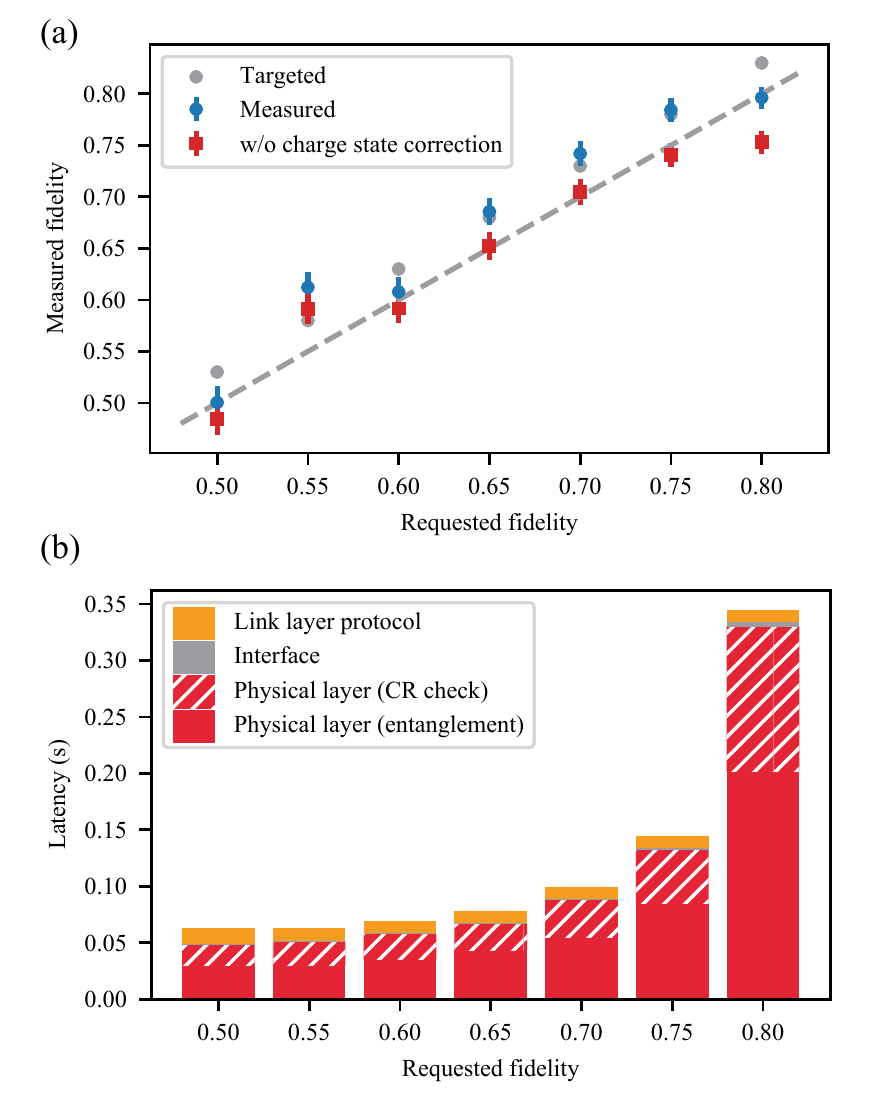}
    \caption{Performance of the entanglement delivery service.
    \textbf{(a)} Measured fidelity of the states delivered by the link layer for varying requested fidelity. Targeted fidelity at the physical layer is 0.03 higher than the link layer protocol's \emph{minimum fidelity} request. When not correcting for wrong charge state events, fidelity is reduced by a few percents (see \cref{sec:link_suppl_m_corr}). Error bars represent \num{1} s.d.
    \textbf{(b)} Average latency of the entanglement delivery per requested fidelity, broken down into sources of latency. Entanglement generation and charge and resonance check at the physical layer are the largest sources of latency (at higher fidelities, more entanglement attempts are required before success). Running the link layer protocol introduces a small but measurable overhead (\SI{\approx 10}{ms}) to the entanglement generation procedure, which does not depend on the requested fidelity, and that could be mitigated by requesting multiple entangled states in a single instruction. The communication delays between quantum network controller and quantum device controller (Interface) introduce negligible overall latency.}
    \label{fig:link_figure_4}
\end{figure}

\subsection{Latency vs Fidelity}

The QEGP interface allows its user to request entangled pairs at various minimum fidelities.
For physical reasons, higher fidelities will result in lower entanglement generation rates~\cite{stockill_phase-tuned_2017, humphreys_deterministic_2018}.
The trade-off between fidelity and throughput is particularly interesting in a scenario where some applications might require high-fidelity entangled pairs and are willing to wait a longer time, while others might prefer lower-fidelity states but higher rates~\cite{dahlberg_link_2019}.
Clearly, for the link layer to offer a range of fidelities to choose from, the underlying physical layer must support such a range.
We benchmark the capabilities of the link layer and of the physical layer to deliver states at various fidelities in a single application by measuring the $\braket{\mathrm{XX}}$, $\braket{\mathrm{YY}}$ and $\braket{\mathrm{ZZ}}$ correlators (and their $\pm$ variations, as we did above, for a total of $3 \times 4 = 12$ correlators) for seven different target fidelities, (\num{0.50}, \num{0.55}, \num{0.60}, \num{0.65}, \num{0.70}, \num{0.75}, \num{0.80}).
We generate \num{1500} entangled states per fidelity, for a total of \num{10500} delivered states (see \cref{lst:link_fid} in \cref{sec:link_suppl_apps}).
With this case study, we analyze both the resulting fidelity and the system's latency for different requested fidelities.

The results for measured fidelity versus requested fidelity are shown in \Cref{fig:link_figure_4}a.
It is worth noting that the application iterates over the range of fidelities in real-time, and thus the physical layer is prepared to deliver any of them at any point.
We calibrate the physical layer to deliver states of slightly higher fidelity than the requested ones (\num{0.03} more), since entanglement requests specify the \emph{minimum} desired fidelity.
The measured fidelities are---within measurement uncertainty---always matching or exceeding the requested minimum ones (the dashed gray line in \cref{fig:link_figure_4}a is the $y=x$ diagonal).
As in the previous application, measurement outcomes are post-processed to eliminate tomography errors and events in which the physical devices were in the incorrect charge state (we refer to the latter as \emph{charge state correction}).
For arbitrary applications that use the delivered entangled states for something other than statistical measurements, applying the second correction directly at the link layer might prove challenging, since the information concerning whether to discard an entangled pair is only available at the physical layer \emph{after} the entangled state is delivered to the link layer (when the next CR check is performed).
However, a mechanism to identify \emph{bad} entangled pairs retroactively at the link layer---like the expiry functionality included in the original design of QEGP~\cite{dahlberg_link_2019}---could be used to discard entangled states after they have been delivered by the physical layer.
For completeness, we also report, again in \Cref{fig:link_figure_4}a, the measured fidelity when the wrong charge state correction is not applied.

For each requested fidelity we also measure the entanglement generation latency~\cite{dahlberg_link_2019}, defined as the time between the issuing of the \texttt{CREATE} request to the link layer, until the successful entanglement outcome reported by the physical layer (refer to \cref{fig:link_figure_3}a for a diagram of the events in between these two).
\Cref{fig:link_figure_4}b shows the measured average latency, grouped by requested fidelity and broken down into the various sources of latency.
When calculating the average latencies, we have ignored entanglement requests that required more than \SI{10}{s} to be fulfilled. These high-latency requests correspond to the horizontal plateaus of \Cref{fig:link_figure_3}d (see \cref{sec:link_suppl_off_resonant} for details).
The main contribution to the total latency comes from the entanglement generation process at the physical layer, followed by the NV center preparation time (CR check).
Both latency values are consistent with the expected number of entanglement attempts required by the single-photon entanglement protocol employed at the physical layer~\cite{humphreys_deterministic_2018}.
The link layer protocol adds, on average, \SI{\approx 10}{ms} of extra latency to all requests, regardless of their fidelity.
This is due partly to the synchronization of the \texttt{CREATE} request between the two nodes (i.e.~a simple TCP message), but mostly to the nodes having to wait for the next time bin in the network schedule to start (the larger the time bins, the larger the worst-case waiting time, see \cref{sec:link_suppl_tdma}).
We remark that, by requesting multiple entangled states in a single \texttt{CREATE}, one can distribute this overhead over many generated pairs, to the point where it becomes negligible.
While our applications did not issue multi-pair \texttt{CREATE} requests, this would be the more natural choice for real applications, and would result in better utilization of the allocated time bins.
Finally, the overhead incurred by the interface between microcontrollers is rather small (barely visible in \cref{fig:link_figure_4}b), but could however be further reduced by integrating device controller and network controller into a single device.
It is worth mentioning that, in our simple scenario in which each entanglement request is only submitted to QEGP after the previous one completes, and thus the request queue never grows larger than one element, throughput happens to be almost exactly the same as the inverse of latency, and hence it is not reported here.

Overall, we observe that the extra entanglement generation latency incurred when deploying an abstraction layer (QEGP) on top of the physical layer, while not too modest, is only a small part of the whole, particularly at higher fidelities.
Nevertheless, optimizing the length of TDMA time bins could result in an even smaller overhead (see \cref{sec:link_suppl_tdma}).

\begin{figure}
    \centering
    \includegraphics{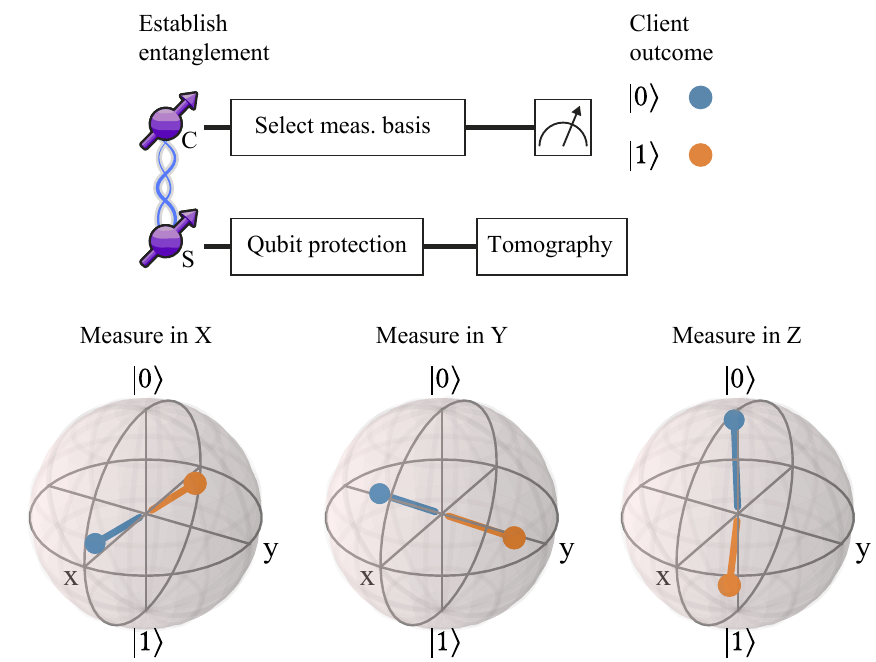}
    \caption{Tomography of states prepared on the server by the remote client. For each chosen measurement axis of the client (X, Y, Z), and for each obtained measurement outcome at the client ($\ket{0}$, $\ket{1}$), a different state is prepared on the server. Plotted on the Bloch spheres are the results of the tomography on the server's qubit. Uncertainties on each coordinate are \num{\approx 0.05} (see \cref{sec:link_suppl_m_corr}). We find an average preparation fidelity of $F =$ \num{0.853(8)}.}
    \label{fig:link_figure_5}
\end{figure}

\subsection{Remote State Preparation}

One of the use cases of the QEGP service is to prepare quantum states on a remote quantum server~\cite{dahlberg_link_2019}.
Remote state preparation is a fundamental step to execute a blind quantum computation application~\cite{broadbent_universal_2009}, whereby a client quantum computer with limited resources can run quantum applications on a powerful remote quantum server using the many qubits the server has, while keeping the performed computation private.

Remote state preparation is different from the previous two cases in that the client can measure its end of the entangled pair as soon as the pair is generated, while the server has to keep its qubit alive waiting for further instructions.
For such a scenario, the client can make use of QEGP's service to issue \textit{R}-type entanglement requests, so that the local end of the entangled pair can be measured (in a certain basis) as soon as it is generated, while the server's qubit can be protected for later usage.
An \textit{R}-type entanglement request results in an \texttt{ENM} command on the client and an \texttt{ENT} command on the server.
For this type of requests (as well as for \textit{M}-type ones), since the local end of the pair is measured immediately, the client's QEGP can skip the Pauli correction used to always deliver $\ket{\Phi^+}$, and can instead apply a classical correction to the received measurement outcome (refer to \cref{sec:link_suppl_pauli_corr}).

To showcase this feature of QEGP we use the client node to prepare the six cardinal states on the server ($\ket{\pm x}$, $\ket{\pm y}$, $\ket{0}$ and $\ket{1}$) by having the client measure its share of the entangled state in the six cardinal bases.
We then let the server measure the prepared states---again in the six cardinal bases---to perform tomography.
For each client measurement basis, and for each server tomography basis, we deliver \num{125} entangled states at a requested fidelity of \num{0.80}, for a total of $6 \times 6 \times 125 = 4500$ remote state preparations (see \cref{lst:link_rsp} in \cref{sec:link_suppl_apps}).
The results are presented in \Cref{fig:link_figure_5}, which displays the tomography of the prepared states on the server, for the three different measurement axes of the client and the two possible measurement outcomes of the client.
The prepared states are affected by the measurement error of the client ($F_0 =$ \num{0.928(3)}, $F_1 = $ \num{0.997(1)}): an error in the measurement of the client's qubit results in an incorrect identification of the state prepared on the server.
By alternating between positive and negative readout orientations, we make sure that the errors affect all prepared states equally, instead of biasing the result.
We note that we exclude, once again, events in which at least one of the two devices was in the wrong charge state, and we correct for the known tomography error on the server (results without corrections are in \cref{sec:link_suppl_m_corr}).
Overall, we find an average remote state preparation fidelity of $F =$ \num{0.853(8)}.
The asymmetry in the fidelity of the $\ket{0}$ and $\ket{1}$ states is caused by the asymmetry in the populations $\braket{01|\ \rho\ |01}$ vs $\braket{10|\ \rho\ |10}$ of the delivered entangled state, which in turn is due to the double $\ket{0}$ occupancy error of the single-photon protocol used to generate entanglement~\cite{humphreys_deterministic_2018, pompili_realization_2021}.
\section{Conclusions}

In summary, we have demonstrated the operation of a link layer and a physical layer for entanglement-based quantum networks.
The link layer abstracts the entanglement generation procedure provided by the physical layer---implemented here with two NV center-based quantum network nodes---into a robust platform-independent service that can be used to run quantum networking applications.
We performed full quantum state tomography of the states delivered by the link layer, tested its ability to deliver states at different fidelities in real-time, and verified remote state preparation of a qubit from the client on the server, a fundamental step towards blind quantum computation~\cite{broadbent_universal_2009}.
We have shown that our implementation of link and physical layers can deliver entangled states at the fidelity requested by the user, despite some marginal inefficiencies---some of which can be addressed in a future version of the protocols (e.g.~avoiding Pauli corrections unless necessary).
We have also quantified the additional latency incurred by deploying the link layer protocol on top of the physical layer.
Although not detrimental, the extra overhead is still noticeable, but can also be scaled down by optimizing the scheduling of entanglement generation requests.
We also acknowledge that scheduling a quantum node's resources is still an open problem~\cite{skrzypczyk_dynamic_2021, vardoyan_stochastic_2021, vardoyan_capacity_2021}, and that the simple TDMA approach taken here might be a suboptimal choice in more advanced quantum networks.

The adoption of the techniques presented here (which are not specific to our diamond devices) by other quantum network platforms~\cite{ritter_elementary_2012, stockill_phase-tuned_2017, rose_observation_2018, nguyen_quantum_2019, stephenson_high-rate_2020, trusheim_transform-limited_2020, son_developing_2020} will boost the development towards large-scale and heterogeneous quantum networks.
Real-time control of memory qubits, as well as the availability of multi-node networks and dynamic network schedules, will enable demonstrations of the higher layers of the network stack~\cite{kozlowski_towards_2019}, which in turn will open the door to end-to-end connectivity on a platform-independent quantum network.

\begin{acknowledgments}
We thank Joris van Rantwijk, Sidney Cadot, Ludo Visser and Nicolas Demetriou for experimental support, Nico Seidler and \"Onder Karpat for useful discussions, and Simon Baier for comments on an earlier version of this manuscript.

We acknowledge financial support from the EU Flagship on Quantum Technologies through the project Quantum Internet Alliance (EU Horizon 2020, grant agreement no.~820445); from the Netherlands Organisation for Scientific Research (NWO) through the Zwaartekracht program Quantum Software Consortium (project no.~024.003.037/3368); from the European Research Council (ERC) through a Consolidator Grant (grant agreement no.~772627 to R.H.)~under the European Union’s Horizon 2020 Research and Innovation Programme.

The datasets that support this manuscript and the software to analyze them are available at Ref.~\cite{link_data_2021}.
\end{acknowledgments}


\bibliographystyle{bib_style}
\bibliography{bibliography}
\onecolumngrid

\clearpage
\begin{center}
\textbf{\large Supplementary Material}
\end{center}
\makeatletter
\renewcommand{\theequation}{S\arabic{equation}}
\renewcommand{\thefigure}{S\arabic{figure}}
\renewcommand{\thetable}{S\arabic{table}}
\renewcommand{\thesection}{S-\Roman{section}}
\makeatother

\setcounter{equation}{0}
\setcounter{figure}{0}
\setcounter{table}{0}
\setcounter{section}{0}

\section{Pre-computed TDMA schedule}
\label{sec:link_suppl_tdma}

As mentioned in \Cref{sec:link_layer}, TDMA network schedules are redundant in a one-link network.
Time-binning network activity also forces nodes to only process entanglement requests at the beginning of a time division, thus introducing latency and idle time.
Particularly, longer time bins potentially result in entanglement requests to wait longer to be processed.
However, an application asking for multiple entangled pairs with just one request would experience smaller average latencies, as all pairs---but the first one---would be generated in close succession.

In our experiments, TDMA schedules are just a constant division of \SI{20}{ms} time bins, each of which is reserved to the only application running.
We chose the duration of the time-bin---somewhat arbitrarily, given the small effect on our experiments---to be equal to \num{1000} communication cycles between the device controller and the network controller (\SI{20}{ms} $= 1000~\times$ \SI{20}{\micro s}).

\section{Single qubit gates implemented}
\label{sec:link_suppl_sqg}
At the physical layer, we implement real-time rotations around the X and Y axes of the qubit Bloch sphere, using a resolution of $\pi/16=$\ang{11.25}. That is, the link layer can request any rotation that is a multiple of $\pi/16$ around either the X or Y axis. The different rotations are performed using Hermite-shaped pulses (as described in Ref.~\cite{pompili_realization_2021}) of calibrated amplitude. The choice of X(Y) rotation axis is implemented using the I(Q) channel of the microwave vector source.

While supported at the link layer, our physical layer currently does not implement Z axis rotations.
Such rotations around the Z axis could be implemented by virtual rotations of the Bloch sphere: a $\pi$ pulse around the Z axis is equivalent to multiplying future I and Q voltages by $-1$. By keeping track of the accumulated Z rotations, and by adjusting I and Q mixing accordingly, one can perform effective Z rotations with very high resolution and virtually no infidelity. The AWGs currently in use have the required capabilities, and the implementation of said Z gates is planned for the near future.

\section{Clock sharing and AWG triggering over longer distances}
\label{sec:link_suppl_timing}
One of the technical challenges of realizing a large scale quantum network is synchronizing equipment at the physical layer across nodes. The synchronization is required to generate entanglement---the photons from the two nodes need to arrive at the same time at the heralding station (compared to their duration, \SI{12}{ns} for NV centers in bulk diamond samples); failing to do so would reduce (or even remove) their indistinguishability, which is required to establish long-distance entanglement~\cite{pompili_realization_2021}.
Our two nodes are located in a single laboratory, on the same optical table, approximately \SI{2}{m} apart.
This allows for some simplifications, for the purpose of demonstrating entanglement delivery using a network stack, which would not be possible over longer distances.
Specifically:
\begin{enumerate}
    \item We use a single laser---the client's---to excite both nodes, as in Ref.~\cite{pompili_realization_2021}. Over longer distances, one would need to phase-lock the excitation lasers at the two nodes to ensure phase-stability of the entangled states.
    
    \item The Device Controllers (ADwin Pro II microcontroller) are triggered every \SI{1}{\us} by the same signal generator, advancing the state machine algorithm that implements the physical layer. This ensures that the two microcontrollers have a common shared clock. Over longer distances, one could use existing protocols (and commercially-available hardware) to obtain a shared clock~\cite{serrano_white_2013}, and use that to trigger the microcontrollers.
    
    \item The two AWGs need to be triggered to play entanglement attempts. In our implementation, one device controller---the server's---triggers both AWGs. This ensures that the triggering delay between the two AWGs is constant, and we can therefore calibrate it out.
    Triggering the AWGs with two independent microcontrollers would result in jitter (realistically on the order of nanoseconds).
    Over larger distances, one could derive---from the shared clock---a periodic trigger signal that is gated by the microcontroller at each node.
    In this way the microcontroller can decide whether the AWG will be triggered on the next cycle, but the accuracy of the trigger's timing will be derived from the shared clock between the nodes, rather than from the microprocessor.
    
    \item The phase stabilization scheme we use, developed in Ref.~\cite{pompili_realization_2021}, is designed to work at a single optical frequency (in our case, the \SI{637}{\nm} emission frequency of the NV center). Over longer distances, conversion of the NV center photons to the telecom band will be necessary to overcome photon loss. The phase stabilization scheme will therefore need to be adapted to new optical frequencies used.
\end{enumerate}
For reference, our client (server) is based on node Charlie (Bob) of the multi-node quantum network presented in Ref.~\cite{pompili_realization_2021}.

\section{Applications}
\label{sec:link_suppl_apps}
Following are pseudocode sequences that describe the applications executed via the quantum network stack. Their purpose is to outline how the applications were executed (sweep order), and the differences between the three applications.

\begin{minipage}{0.7\linewidth}
\begin{lstlisting}[caption=Full quantum state tomography., texcl, label=lst:link_fst]
# The common list of correlators that are going to be measured (client, server).
correlator_list = [
    (-X, -X), (-X, -Y), (-X, -Z), (-X, +X), (-X, +Y), (-X, +Z), 
    (-Y, -X), (-Y, -Y), (-Y, -Z), (-Y, +X), (-Y, +Y), (-Y, +Z), 
    (-Z, -X), (-Z, -Y), (-Z, -Z), (-Z, +X), (-Z, +Y), (-Z, +Z),
    (+X, -X), (+X, -Y), (+X, -Z), (+X, +X), (+X, +Y), (+X, +Z), 
    (+Y, -X), (+Y, -Y), (+Y, -Z), (+Y, +X), (+Y, +Y), (+Y, +Z), 
    (+Z, -X), (+Z, -Y), (+Z, -Z), (+Z, +X), (+Z, +Y), (+Z, +Z)]

def client_application():
    for rep in range(125):
        for corr in correlator_list:
            client_basis = corr[0]
            # Establish the entangled state.
            client_qubit = create_ent(
                with=Server,
                req_type=Keep,
                min_fidelity=0.8)
            # Perform the required rotation.
            client_qubit.rotate_basis(client_basis)
            # Measure the qubit, store the result.
            outcomes[rep, corr] = client_qubit.measure()

def server_application():
    for rep in range(125):
        for corr in correlator_list:
            server_basis = corr[1]
            # Establish the entangled state.
            server_qubit = receive_ent(
                with=Client,
                req_type=Keep,
                min_fidelity=0.8)
            # Perform the required rotation.
            server_qubit.rotate_basis(server_basis)
            # Measure the qubit, store the result.
            outcomes[rep, corr] = server_qubit.measure()
\end{lstlisting}
\end{minipage}

\begin{minipage}{0.7\linewidth}
\begin{lstlisting}[caption=Delivery of entangled states at varying fidelity and rate., texcl, label=lst:link_fid]
# The common list of correlators that are going to be measured (client, server).
correlator_list = [
    (-X, -X), (-X, +X), (-Y, -Y), (-Y, +Y), (-Z, -Z), (-Z, +Z),
    (+X, -X), (+X, +X), (+Y, -Y), (+Y, +Y), (+Z, -Z), (+Z, +Z)]

# The target fidelities to generate.
fidelity_list = [0.50, 0.55, 0.60, 0.65, 0.70, 0.75, 0.80]

def client_application():
    for rep in range(125):
      for fid in fidelity_list:
        for corr in correlator_list:
          client_basis = corr[0]
          # Establish the entangled state.
          client_qubit = create_ent(
                with=Server,
                req_type=Keep,
                min_fidelity=fid)
          # Perform the required rotation.
          client_qubit.rotate_basis(client_basis)
          # Measure the qubit, store the result.
          outcomes[rep, fid, corr] = client_qubit.measure()

def server_application():
    for rep in range(125):
      for fid in fidelity_list:
        for corr in correlator_list:
          server_basis = corr[1]
          # Establish the entangled state.
          server_qubit = receive_ent(
                with=Client,
                req_type=Keep,
                min_fidelity=fid)
          # Perform the required rotation.
          server_qubit.rotate_basis(server_basis)
          # Measure the qubit, store the result.
          outcomes[rep, fid, corr] = server_qubit.measure()
\end{lstlisting}
\end{minipage}

\begin{minipage}{0.7\linewidth}
\begin{lstlisting}[caption=Remote preparation of a qubit on the server., texcl, label=lst:link_rsp]
# The common list of correlators that are going to be measured (client, server).
correlator_list = [
    (-X, -X), (-X, -Y), (-X, -Z), (-X, +X), (-X, +Y), (-X, +Z), 
    (-Y, -X), (-Y, -Y), (-Y, -Z), (-Y, +X), (-Y, +Y), (-Y, +Z), 
    (-Z, -X), (-Z, -Y), (-Z, -Z), (-Z, +X), (-Z, +Y), (-Z, +Z),
    (+X, -X), (+X, -Y), (+X, -Z), (+X, +X), (+X, +Y), (+X, +Z), 
    (+Y, -X), (+Y, -Y), (+Y, -Z), (+Y, +X), (+Y, +Y), (+Y, +Z), 
    (+Z, -X), (+Z, -Y), (+Z, -Z), (+Z, +X), (+Z, +Y), (+Z, +Z)]

def client_application():
    for rep in range(125):
        for corr in correlator_list:
            client_basis = corr[0]
            # Establish the entangled state and measure in the specified basis.
            outcomes[rep, corr] = create_ent(
                with=Server,
                req_type=RemoteStatePreparaion,
                measurement_basis=client_basis,
                min_fidelity=0.8)

def server_application():
    for rep in range(125):
        for corr in correlator_list:
            server_basis = corr[1]
            # Establish the entangled state.
            server_qubit = receive_ent(
                with=Client,
                req_type=RemoteStatePreparation,
                min_fidelity=0.8)
            # Perform the required rotation.
            server_qubit.rotate_basis(server_basis)
            # Measure the qubit, store the result.
            outcomes[rep, corr] = server_qubit.measure()
\end{lstlisting}
\end{minipage}

\section{Post-measurement Pauli correction}
\label{sec:link_suppl_pauli_corr}
The physical layer, depending on the specific quantum platform, will deliver in general one of the four possible Bell states. With the single photon protocol we employ, the physical layer can produce either  $\ket{\Psi^+} = (\ket{01} + \ket{10})/\sqrt{2}$ or $\ket{\Psi^-} = (\ket{01} - \ket{1})/\sqrt{2}$, see Refs.~\cite{humphreys_deterministic_2018, pompili_realization_2021}.

We choose to offer the generation of $\ket{\Phi^+} = (\ket{00} + \ket{11})/\sqrt{2}$ as the link layer service (the choice of the specific state is arbitrary).
In principle, one could also make the generated state a parameter of the link layer request.
To deliver the desired Bell state, the link layer applies, for \textit{K}-type requests (entangle and keep), a Pauli correction, by requesting a $\pi$ rotation around either the X or Y axis on the client's qubit.
For \textit{M}-type (entangle and measure) and \textit{R}-type (remote state preparation) requests, the physical layer performs a measurement in a basis prespecified---using the \texttt{PMG} command---and reports the measurement outcome, as well as which state was generated, to the link layer.
The link layer can, depending on the generated state and on the chosen measurement basis, apply a classical bit flip on the client's outcome to obtain the correct measurement statistics.
In particular, the link layer flips the measurement outcome of the client in the following cases: 
\begin{enumerate*}[label=(\alph*)]
    \item State delivered $\ket{\Psi^+}$, measurement basis $\pm Y$;
    \item State delivered $\ket{\Psi^-}$, measurement basis $\pm X$;
    \item State delivered $\ket{\Psi^+}$ or $\ket{\Psi^-}$, measurement basis $\pm Z$.
\end{enumerate*}

\section{Results with and without corrections}
\label{sec:link_suppl_m_corr}
The data presented in the main text is corrected for known measurement errors, and events in which at least one of the two devices was in the wrong charge state are removed (the CR check following the delivery of entanglement reports zero counts). While it is useful to correct for such errors in order to obtain the most faithful reconstruction of the delivered states, these errors cannot always be avoided in a real network scenario. 
For completeness, we report here the same results as in \Cref{sec:link_eval}, first without any corrections applied, and then with only the measurement error correction applied. All the results, the raw datasets, and the software to analyze them, are available at Ref.~\cite{link_data_2021}.

\subsection{Full Quantum State Tomography}

The events in which the two devices generated \num{0} photon counts in the following CR check were \num{37} for the client and \num{380} for the server (out of the \num{4500} total). When combined, (client or server in the wrong charge state), we obtain \num{417} events (in zero events both client and server were in the wrong charge state).
Without any corrections (tomography errors or wrong charge state), we obtain the following density matrix (which has a fidelity with the target Bell state F=\num{0.681(16)}):
\[
\mathrm{Re}[\rho] = \begin{pmatrix}
  0.397(9) & 0.011(9) & 0.001(7) & 0.256(14)\\
  0.011(9) & 0.058(14) & -0.005(13) & -0.007(9)\\
  0.001(7) & -0.005(13) & 0.092(12) & -0.027(13)\\
  0.256(14) & -0.007(9) & -0.027(13) & 0.452(9)\\
\end{pmatrix},
\]
\[
\mathrm{Im}[\rho] = \begin{pmatrix}
  0 & 0.000(18) & -0.029(9) & 0.036(9)\\
  -0.000(18) & 0 & 0.010(12) & -0.002(8)\\
  0.029(9) & -0.010(12) & 0 & -0.000(8)\\
  -0.036(9) & 0.002(8) & 0.000(8) & 0\\
\end{pmatrix}
\]

Only applying tomography error correction (but not removal of wrong charge state events) yields the following density matrix (fidelity F=\num{0.744(11)}):
\[
\mathrm{Re}[\rho] = \begin{pmatrix}
  0.421(7) & -0.001(4) & -0.013(5) & 0.300(8)\\
  -0.001(4) & 0.022(8) & -0.020(6) & -0.021(7)\\
  -0.013(5) & -0.020(6) & 0.091(5) & -0.015(5)\\
  0.300(8) & -0.021(7) & -0.015(5) & 0.466(5)\\
\end{pmatrix}
\]
\[
\mathrm{Im}[\rho] = \begin{pmatrix}
  0 & 0.004(4) & -0.018(3) & 0.032(6)\\
  -0.004(4) & 0 & 0.021(6) & 0.002(5)\\
  0.018(3) & -0.021(6) & 0 & 0.002(5)\\
  -0.032(6) & -0.002(5) & -0.002(5) & 0\\
\end{pmatrix}
\]

\subsection{Fidelity vs Rate}

The events in which the two devices generated \num{0} photon counts in the following CR check were \num{74} for the client and \num{709} for the server (out of the \num{10500} total). When combined, (client or server in the wrong charge state), we obtain \num{781} events (there were two events in which both client and server were in the wrong charge state).
Without any corrections (tomography errors or wrong charge state), we obtain the following delivered fidelities: \num{0.454(18)}, \num{0.540(18)}, \num{0.548(17)}, \num{0.596(17)}, \num{0.640(16)}, \num{0.674(16)}, \num{0.679(15)}.
Only applying tomography error correction (but not removal of wrong charge state events) yields the following fidelities: \num{0.485(15)}, \num{0.591(14)}, \num{0.592(14)}, \num{0.652(13)}, \num{0.705(13)}, \num{0.741(12)}, \num{0.753(11)}.

\subsection{Remote State Preparation}

As mentioned in the main text, for the remote state preparation analysis, we only apply the tomography error correction for the server, while remove wrong charge state events of both the server and the client. The events in which the two devices generated \num{0} photon counts in the following CR check were \num{29} for the client and \num{365} for the server (out of the \num{4500} total). When combined, (client or server in the wrong charge state), we obtain \num{394} events (there were zero events in which both client and server were in the wrong charge state).
Following are the numerical values that result in the plot in the main text (average fidelity F=\num{0.853(8)}):
\begin{table}[H]
    \centering
    \begin{tabular}{|p{2.5cm}|ccc|c|}
    \hline
    Client & \multicolumn{4}{c|}{Server}\\
    & $\braket{\mathrm{X}}$ & $\braket{\mathrm{Y}}$ & $\braket{\mathrm{Z}}$ & Fidelity\\ \hline
    Measured $\ket{+X}$ & \num{0.634(48)} & \num{-0.123(62)} & \num{-0.004(59)} & \num{0.817(24)}\\
    Measured $\ket{+Y}$ & \num{-0.028(58)} & \num{-0.650(45)} & \num{0.005(61)} & \num{0.825(23)}\\
    Measured $\ket{+Z}$ & \num{-0.081(65)} & \num{-0.083(66)} & \num{0.849(31)} & \num{0.924(16)}\\
    Measured $\ket{-X}$ & \num{-0.645(43)} & \num{0.135(59)} & \num{0.030(63)} & \num{0.823(22)}\\
    Measured $\ket{-Y}$ & \num{0.026(65)} & \num{0.719(40)} & \num{-0.013(61)} & \num{0.860(20)}\\
    Measured $\ket{-Z}$ & \num{0.032(58)} & \num{-0.069(58)} & \num{-0.736(39)} & \num{0.868(19)}\\
    \hline
    \end{tabular}
\end{table}

Without any corrections (tomography errors or wrong charge state), we obtain the following prepared states, with average fidelity F=\num{0.807(10)}:
\begin{table}[H]
    \centering
    \begin{tabular}{|p{2.5cm}|ccc|c|}
    \hline
    Client & \multicolumn{4}{c|}{Server}\\
    & $\braket{\mathrm{X}}$ & $\braket{\mathrm{Y}}$ & $\braket{\mathrm{Z}}$ & Fidelity\\ \hline
    Measured $\ket{x}$ & \num{0.534(55)} & \num{-0.090(62)} & \num{0.009(62)} & \num{0.767(27)}\\
    Measured $\ket{y}$ & \num{0.024(60)} & \num{-0.582(51)} & \num{-0.013(62)} & \num{0.791(26)}\\
    Measured $\ket{0}$ & \num{-0.073(69)} & \num{-0.072(69)} & \num{0.786(42)} & \num{0.893(21)}\\
    Measured $\ket{-x}$ & \num{-0.552(49)} & \num{0.143(61)} & \num{0.055(63)} & \num{0.776(24)}\\
    Measured $\ket{-y}$ & \num{0.052(64)} & \num{0.623(47)} & \num{-0.018(62)} & \num{0.811(23)}\\
    Measured $\ket{1}$ & \num{0.030(57)} & \num{-0.028(55)} & \num{-0.606(46)} & \num{0.803(23)}\\
    \hline
    \end{tabular}
\end{table}

When only applying tomography error correction, we find an average preparation fidelity F=\num{0.829(9)}:
\begin{table}[H]
    \centering
    \begin{tabular}{|p{2.5cm}|ccc|c|}
    \hline
    Client & \multicolumn{4}{c|}{Server}\\
    & $\braket{\mathrm{X}}$ & $\braket{\mathrm{Y}}$ & $\braket{\mathrm{Z}}$ & Fidelity\\ \hline
    Measured $\ket{x}$ & \num{0.573(49)} & \num{-0.096(59)} & \num{0.010(58)} & \num{0.786(24)}\\
    Measured $\ket{y}$ & \num{0.025(56)} & \num{-0.624(45)} & \num{-0.014(59)} & \num{0.812(23)}\\
    Measured $\ket{0}$ & \num{-0.078(64)} & \num{-0.077(65)} & \num{0.843(32)} & \num{0.921(16)}\\
    Measured $\ket{-x}$ & \num{-0.592(44)} & \num{0.153(57)} & \num{0.059(59)} & \num{0.796(22)}\\
    Measured $\ket{-y}$ & \num{0.056(61)} & \num{0.667(41)} & \num{-0.020(59)} & \num{0.834(20)}\\
    Measured $\ket{1}$ & \num{0.032(54)} & \num{-0.030(53)} & \num{-0.650(40)} & \num{0.825(20)}\\
    \hline
    \end{tabular}
\end{table}

\section{NV center resonance control}
\label{sec:link_suppl_off_resonant}
The two quantum network nodes use different techniques to control the resonance of their NV centers (see Ref.~\cite{pompili_realization_2021} for implementations details).
The server uses an off-resonant charge randomization strategy: when its NV center is not on resonance (it does not pass the charge and resonance check), it can apply an off-resonant (green, \SI{515}{\nm}) laser pulse to shuffle the charge environment and probabilistically recover the correct charge and resonance state. The server cannot get \emph{stuck} in a non-resonance state: in a few tens of failed CR checks and green laser pulses (overall less than \SI{1}{\ms}) the NV center will be in resonance again.

The client, which needs to be tuned in resonance with the other node, uses a resonant strategy. When in the wrong charge state (zero counts during CR check), it applies a resonant laser pulse (yellow, \SI{575}{nm}, NV${}^0$ zero-phonon line) to go back to NV${}^-$.
To bring NV${}^-$ in resonance with the necessary lasers, it adjusts a biasing voltage applied to the diamond sample, which shifts the resonance frequencies. This process is mostly automated. However, occasional human intervention is still required when the resonance frequencies of the NV center shift too far---for example due to a charge in the vicinity of the NV center changing position in the lattice---for the automatic mechanism to find its way back.
The horizontal steps in \Cref{fig:link_figure_3}d are due to the jumps in the client's NV optical transitions, which then require manual optimization of the laser frequencies and/or the diamond biasing voltage---depending on the magnitude of the frequency shift, it requires tens of seconds to a few minutes to recover the optimal resonance condition.

\begin{table}[]
\centering
\caption{List of possible outcomes of the physical layer to link layer requests.}
\label{tab:link_possible_outcomes}
\begin{tabular}{|p{1.8cm}|l|p{8cm}|}
\hline
Link layer request & Physical layer outcome & Description\\
\hline
\texttt{INI} & \texttt{SUCCESS} & Qubit initialization is always successful.\\ 
\hline
\texttt{MSR} & \texttt{SUCCESS\_0} & Measurement outcome is $\ket 0$.\\
& \texttt{SUCCESS\_1} & Measurement outcome is $\ket 1$.\\
\hline
\texttt{SQG} & \texttt{SUCCESS} & Single qubit gates are always successful.\\
\hline
\texttt{PMG} & \texttt{SUCCESS} & Updating the pre-measurement gate information is always successful.\\
\hline
\texttt{ENT} & \texttt{SUCCESS\_PSI\_PLUS} & Entanglement generation was successful, the state generated was $\ket{\Psi^+}$.\\
& \texttt{SUCCESS\_PSI\_MINUS} & Entanglement generation was successful, the state generated was $\ket{\Psi^-}$.\\
\hline
\texttt{ENM} & \texttt{SUCCESS\_PSI\_PLUS\_0} & Entanglement generation was successful, the state generated was $\ket{\Psi^+}$, and the measurement outcome was $\ket 0$.\\
& \texttt{SUCCESS\_PSI\_PLUS\_1} & Entanglement generation was successful, the state generated was $\ket{\Psi^+}$, and the measurement outcome was $\ket 1$.\\
& \texttt{SUCCESS\_PSI\_MINUS\_0} & Entanglement generation was successful, the state generated was $\ket{\Psi^-}$, and the measurement outcome was $\ket 0$.\\
& \texttt{SUCCESS\_PSI\_MINUS\_1} & Entanglement generation was successful, the state generated was $\ket{\Psi^-}$, and the measurement outcome was $\ket 1$.\\
\hline
\texttt{ENT}, \texttt{ENM} & \texttt{ENT\_FAILURE} & Entanglement generation was attempted and failed.\\
& \texttt{ENT\_SYNC\_FAILURE} & Entanglement generation was not attempted because the synchronization step failed (other node is busy).\\
\hline
Any request & \texttt{HARDWARE\_FAILURE} & The node has experienced a hardware problem and cannot fulfill requests.\\
\hline
\end{tabular}
\end{table}

\begin{figure}
    \centering
    \includegraphics{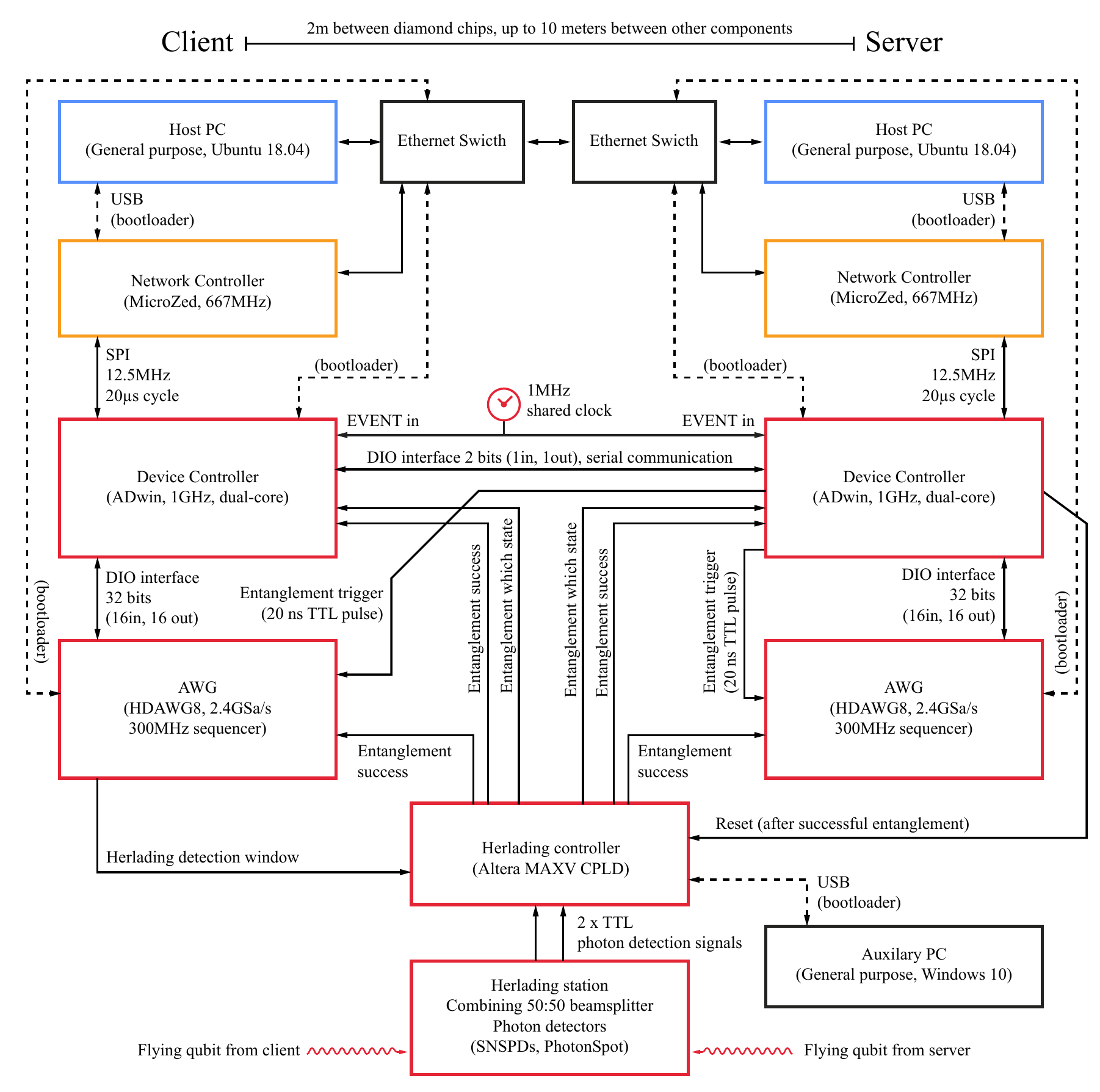}
    \caption{Schematic of the connections between client and server and among the various components of each quantum network node.
    The dashed lines represent connections used when flashing devices (boot loading), and they are not used during the real-time operation of the network stack.
    Not shown are additional optical and electronic components used to control the qubits, see Ref.~\cite{pompili_realization_2021} for details on the equipment.
    }
    \label{fig:link_suppl_setup}
\end{figure}
\clearpage
\glsaddall
\printnoidxglossary[type=\acronymtype, title={List of Abbreviations}, toctitle={}]
\end{document}